\documentclass[aps,prd,amsmath,floats,floatfix,twocolumn,
  superscriptaddress,nofootinbib,showpacs]{revtex4-2}

\usepackage{graphicx}
\setkeys{Gin}{width=3.5in}
\usepackage{tikz}
\usepackage{soul}
\usepackage{listings}
\usepackage{placeins}
\usepackage{amssymb}
\usepackage{tabularx}
\usetikzlibrary{arrows.meta, positioning, shapes.geometric}
\usepackage[export]{adjustbox}
\usepackage{placeins}
\usepackage[caption=false]{subfig}
\usepackage{booktabs}
\usepackage{multirow}
\usepackage{stfloats}
\usepackage{orcidlink}
\usepackage[normalem]{ulem}
\usepackage{hyperref}
\usepackage[capitalise]{cleveref}
\usepackage{etoolbox}
\usepackage{ragged2e}
\usepackage{afterpage}
\usepackage{capt-of}
\usepackage{enumitem}

\crefname{section}{Sec.}{Sec.}
\Crefname{section}{Section}{Sections}
\newcommand{\ccite}[1]{Ref.~\cite{#1}}

\newcommand{\spec}{SpEC}
\newcommand{\spectre}{SpECTRE}

\newcommand{\pomega}{\Omega_0}
\newcommand{\padot}{\dot{a}_0}
\newcommand{\psep}{D_0}
\newcommand{\mnras}{Mon.\ Not.\ R.\ Astron.\ Soc.}

\begin{document}

\title{Data-driven acceleration of eccentricity reduction for binary black hole simulations}

\newcommand{\caltech}{\affiliation{Theoretical Astrophysics, Walter Burke
Institute for Theoretical Physics, California Institute of Technology, Pasadena,
California 91125, USA}}
\newcommand{\uzh}{\affiliation{Department of Astrophysics, University of Zurich, Winterthurerstrasse 190, 8057 Zurich, Switzerland}}
\newcommand{\cornell}{\affiliation{Cornell Center for Astrophysics and Planetary Science, Cornell University, Ithaca, New York 14853, USA}}

\author{Vittoria Tommasini\,\orcidlink{0009-0001-4440-9751}} \email{vtommasini@caltech.edu} \caltech
\author{Nils L. Vu\,\orcidlink{0000-0002-5767-3949}} \email{nilsvu@caltech.edu} \caltech \uzh
\author{Mark A. Scheel\, \orcidlink{0000-0001-6656-9134}}
\email{scheel@tapir.caltech.edu} \caltech
\author{Saul A. Teukolsky\, \orcidlink{0000-0001-9765-4526}}
\email{saul@caltech.edu} \caltech \cornell

\date{April 27, 2026}

\begin{abstract}
Reducing orbital eccentricity in numerical relativity simulations of binary black holes is essential for producing astrophysically relevant gravitational wave models, as many of these systems are expected to be near circular in nature. Standard eccentricity-reduction procedures rely on iterative schemes, often requiring four or more trial simulations to achieve desired thresholds.
This approach is computationally expensive because each trial simulation adds $\mathord{\sim}10\%$ to the total simulation run-time of multiple weeks to months.
We introduce a data-driven approach that accelerates this process by learning the values of the initial orbital frequency, $\Omega_0$, and radial velocity, $\dot{a}_0$, that yield an evolution with small eccentricity. This is done using a Gaussian process regression model trained on an archive of previously eccentricity-reduced numerical relativity simulations. For all configurations tested, using the trained model consistently reduces the number of required eccentricity-reduction iterations to just zero or one, significantly lowering computational costs relative to post-Newtonian initial guesses.
These results demonstrate the power of data-driven methods in accelerating expensive numerical relativity simulations.
\end{abstract}

\maketitle

\section{Introduction}
Numerical relativity (NR) simulations of binary black holes (BBHs) are an essential ingredient in modeling gravitational-wave signals observed by ground-based detectors, such as LIGO \cite{ligo} and Virgo \cite{virgo, Abbott2016}, as well as future space-based missions, such as \textit{LISA} \cite{LISAwhitepaper, Centrella2010}. For the majority of sources targeted by ground-based detectors, the inspiral and merger are expected to proceed in nearly quasicircular orbits \cite{Centrella2010, CentrellaBaker2010, Peters1964, Romero-Shaw2019}. Achieving low-eccentricity configurations is therefore essential for producing astrophysically relevant waveforms and for constructing accurate waveform catalogs. However, selecting initial orbital parameters that yield a desired eccentricity is nontrivial and requires careful control of the initial data \cite{Pfeiffer2007, Buonanno2011, Ramos2019, sxscatalog}. Space-based detectors, such as \textit{LISA}, are expected to also observe systems that retain substantial eccentricity throughout the inspiral \cite{Porter2010, Key2011, LVK}, in which case eccentricity becomes an intrinsic physical parameter \cite{Nee2025, Pankaj2022, Wang2024}. Nevertheless, even in this broader context, the ability to robustly control eccentricity in NR simulations remains a fundamental requirement for accurate waveform modeling \cite{Gamboa2025}.

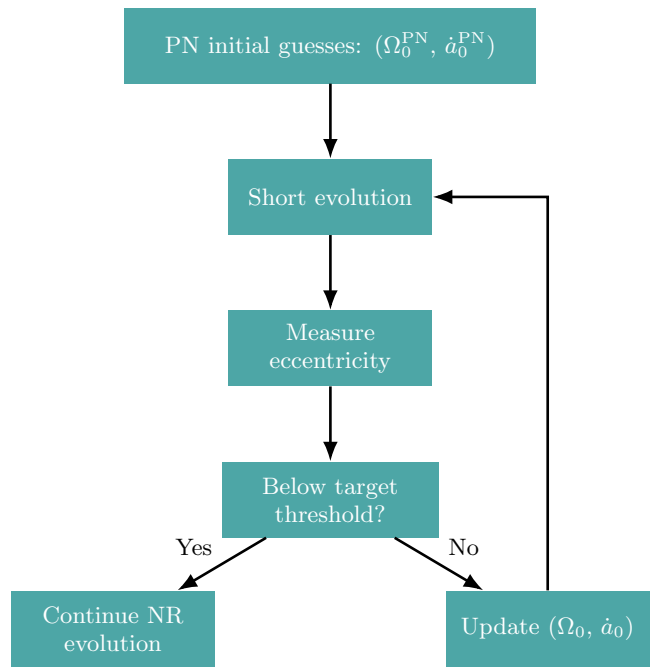
\begin{figure}
\begin{tikzpicture}[
  font=\small,
  node distance=1cm,
  >=Latex,
  box/.style={
    rectangle,
    fill=teal!70,
    text=white,
    align=center,
    text width=0.6\columnwidth,
    minimum height=1cm,
    inner sep=4pt
  },
  midbox/.style={
    rectangle,
    fill=teal!70,
    text=white,
    align=center,
    text width=0.28\columnwidth,
    minimum height=1cm,
    inner sep=4pt
  },
  decision/.style={
    rectangle,
    fill=teal!70,
    text=white,
    align=center,
    text width=0.3\columnwidth,
    minimum height=1cm,
    inner sep=4pt
  },
  arrow/.style={->, line width=1pt},
]
\node (pn) [box] {PN initial guesses: $(\Omega_0^{\rm PN},\, \dot{a}_0^{\rm PN})$};
\node (short) [midbox, below=of pn] {Short evolution};
\node (meas) [midbox, below=of short] {Measure eccentricity};
\node (thresh) [decision, below=of meas] {Below target threshold?};
\node (cont) [midbox, below left=0.7cm and 0.01\columnwidth of thresh] {Continue NR evolution};
\node (update) [midbox, below right=0.7cm and 0.01\columnwidth of thresh] {Update $(\Omega_0,\, \dot{a}_0)$};
\draw[arrow] (pn) -- (short);
\draw[arrow] (short) -- (meas);
\draw[arrow] (meas) -- (thresh);
\draw[arrow] (thresh) -- node[midway, above left] {Yes} (cont);
\draw[arrow] (thresh) -- node[midway, above right] {No} (update);
\draw[arrow] (update.north) |- (short.east);
\end{tikzpicture}
\caption{\justifying Current eccentricity-reduction schemes consist of starting with initial guesses for the orbital parameters, running a short numerical evolution, and measuring the eccentricity from oscillations in $\Omega$ and $\dot{a}$. If the eccentricity is below the target threshold, the simulation continues. If it is not, the orbital parameters are adjusted, and the process is repeated until the eccentricity reaches the chosen threshold. In this work, we focus on improving the first step by utilizing past data to make a better initial guess.}
\label{fig:iterative-scheme}
\end{figure}

In practice, eccentricity is not solely determined by the initial orbital parameters, but is sensitive to the specific gauge conditions, free data choices, and boundary conditions used in the construction of initial data, effectively becoming entangled with gauge effects and junk radiation \cite{Varma2018}. Since eccentricity-reduction procedures operate precisely in this early transient regime \cite{Pfeiffer2007, Buonanno2011, Zhang2013, Habib2025}, determining suitable initial orbital parameters remains a nontrivial task across numerical relativity frameworks \cite{Healy2020, Buchman2012}.

Eccentricity reduction is typically achieved through iterative procedures during which short evolutions are used to measure residual eccentricity and update the initial orbital parameters \cite{Buonanno2011, Buchman2012, Habib2025}. This approach is employed both in generalized-harmonic formulations, with excision-based initial data, and in moving puncture formulations \cite{Buonanno2011, Ramos2019, Habib2021}. Initial guesses for the orbital parameters are typically estimated using post-Newtonian (PN) or effective-one-body approximations. Then a short NR simulation is performed to measure the eccentricity, either from oscillations in the orbital dynamics, or from the emitted gravitational radiation \cite{Buonanno2011, Nee2025, Purrer2012}. The orbital parameters are then adjusted using analytic correction formulas, and the simulation is restarted. This process is repeated as many times as it takes to achieve the target eccentricity. Figure~\ref{fig:iterative-scheme} shows a typical iterative reduction scheme, as explained in more detail in \ccite{Habib2025}. Furthermore, eccentricity reduction becomes increasingly costly as the complexity of the binary system grows. Using runtime measurements from simulations in our Simulating eXtreme Spacetimes (SXS) catalog \cite{sxscatalog}, we find that each eccentricity-reduction iteration adds $
\mathord{\sim}10\%$ computing time to a simulation that takes multiple weeks to months to complete, with typical systems requiring three to four iterations, and complicated systems requiring up to seven iterations.

The problem is further complicated by the ambiguity in the definition of eccentricity itself. The absence of a unique, gauge-invariant notion of separation or phase in fully relativistic binaries means eccentricity does not have a single canonical definition as it does in Newtonian theory \cite{Buonanno2011, Mroue2018, Shaikh2025, Ramos-Buades2022}. Even within post-Newtonian theory, multiple eccentricity parameters are commonly used \cite{Loutrel2018}. As a result, eccentricity in NR is typically inferred indirectly \cite{Buonanno2011, Islam2025, Mroue2018}, and different choices can yield quantitatively different estimates, even for the same simulations \cite{Islam2025, RamosScheel2022, Bonino2024, Healy2020, Mroue2018}. Eccentricity reduction should thus be understood as an operational procedure that is inherently tied to specific gauge choices and diagnostics.

At the same time, large archives of NR simulations now exist in which eccentricity-reduction procedures have already been applied, such as the SXS catalog \cite{sxscatalog}. In this work, we introduce a data-driven method of accelerating the construction of low-eccentricity binary black hole initial data by learning directly from these existing NR simulations. Rather than modifying the definition or measurement of eccentricity, we leverage the outcomes of the current eccentricity-reduction pipeline to predict initial orbital parameters for new configurations. We train a machine learning model to predict corrections to PN initial guesses for the orbital frequency, $\Omega_0$, and radial velocity, $\dot{a}_0$, using these previously converged simulations.

We employ Gaussian process regression (GPR), a nonparametric Bayesian regression method well suited to modeling smooth, multi-dimensional functions, while also providing calibrated uncertainty estimates \cite{Rasmussen2006}. We train the predictive model on residuals between the PN values of orbital parameters that yield zero eccentricity and the values of those parameters obtained from low-eccentricity simulations in the SXS catalog \cite{sxscatalog}. Once trained, our model predicts near-optimal initial orbital parameters for new configurations, substantially reducing the number of eccentricity-reduction iterations required to reach zero eccentricity.

This strategy represents a new approach in using machine learning within NR workflows. Rather than replacing physical models or algorithms altogether, our approach accelerates an already established pipeline by merely providing a better initial guess, effectively leveraging the cost of past simulations to speed up future ones. While machine learning has been successfully applied in adjacent areas of gravitational physics, such as to accelerate numerical subroutines and improve numerical accuracy within NR simulations \cite{Dieselhorst, Helfer}, its use as a tool to directly accelerate the generation of new simulations has received comparatively little attention.

The remainder of this paper is organized as follows: In Sec. ~\ref{sec:methods} we describe our methods, including the setup of BBH initial data in NR simulations and the construction, utilization, and validation of our Gaussian process regression framework. In Sec. ~\ref{sec:results} we present our results, beginning with tests on equal-mass, nonspinning binaries, followed by unequal-mass, nonspinning, and aligned-spin configurations, and finally unequal-mass, precessing-spin binaries; we then validate our method on new simulations, comparing GPR-predicted initial parameters against traditional PN-based eccentricity-reduction procedures, demonstrating significant reductions in the number of required iterations. In Sec. ~\ref{sec:discussion} we discuss the implications of this approach for accelerating NR simulations and potential extensions of this work, concluding our work in Sec. ~\ref{sec:conclusion}.

\section{Methods}
\label{sec:methods}
\subsection{Simulation setup}

NR initial data for BBH simulations are specified in terms of a set of intrinsic and orbital parameters, including the mass ratio $q$, initial coordinate separation, $\psep$, and the individual black hole spin, $\boldsymbol{S_1}$ and $\boldsymbol{S_2}$. In addition, quasicircular initial data are parametrized by the initial orbital frequency, $\Omega_0$, and the initial radial velocity $\dot{a}_0$, which together determine the tangential and radial components of the orbital motion \cite{Mendes}.

The construction of BBH initial data in NR is done by solving the Einstein constraint equations on a spatial hypersurface, subject to a choice of freely specifiable data, gauge conditions, and boundary conditions. In the numerical relativity codes \spec{}~\cite{spec} and \spectre{}~\cite{spectre}, this is accomplished through the use of the extended conformal thin-sandwich formulation, combined with excision boundary conditions on the black hole horizons and asymptotically flat outer boundary conditions \cite{York1999, PfeifferYork2003, CookPfeiffer2004, Pfeiffer2004, BaumgarteShapiro, Vuellpiticsolver}. The freely specifiable data are chosen as a superposition of boosted, spinning Kerr-Schild black holes, resulting in what is called superposed Kerr-Schild (SKS) initial data \cite{Lovelace2008}.

Given an initial guess for $(\Omega_0$, $\dot{a}_0)$, \spec{} obtains low-eccentricity initial data via an iterative eccentricity-reduction procedure \cite{Pfeiffer2007, Buonanno2011}. This procedure consists of evolving the system for a short time, measuring the residual eccentricity from oscillations in orbital quantities, and updating $(\Omega_0,\dot{a}_0)$ accordingly (see \cref{fig:iterative-scheme}). This iterate-measure-correct loop is repeated until the eccentricity falls below a chosen threshold (in this work, $e \lesssim 10^{-3}$). Recent improvements to this procedure have significantly increased its robustness and efficiency \cite{Habib2025}. This work is solely concerned with improving the first step of this procedure, finding better initial guesses.

The initial guess for $(\Omega_0,\dot{a}_0)$ is currently chosen using PN approximations. PN
theory is an expansion of the two-body equations of motion in powers of the orbital velocity $v/c$. For widely separated and slowly moving binaries, PN approximations accurately capture the dynamics of the system. Including higher-order PN terms allows for more relativistic corrections, such as higher-order spin-orbit and spin-spin couplings, radiation-reaction effects, and nonlinear interactions. However, PN theory breaks down as binaries approach merger and relativistic effects become more prominent.

In this work, we employ two PN-based prescriptions to obtain a low-eccentricity initial guess, which we refer to as low-order PN (LOPN) and high-order PN (HOPN). The LOPN prescription corresponds to the approach currently used in \spec{} \cite{KidderPN, Bohe2019}. This approximation has provided initial values for $\Omega_0$ and $\dot{a}_0$ that served as a starting point for most of the simulations in the SXS catalog~\cite{sxscatalog}. The HOPN prescription is a new implementation at higher PN order provided by the \texttt{PostNewtonian.jl} software package \cite{Boyle}. One of our goals is to determine whether utilizing a higher-order initial guess for $(\Omega_0,\dot{a}_0)$ is sufficient to improve eccentricity reduction, without the need for machine learning.

\subsection{Gaussian process regression}
Gaussian process regression is a flexible, nonparametric Bayesian approach to solving regression problems. Instead of assuming a fixed parametric form for the function that one desires to model, a probability distribution is placed over functions. A Gaussian process is fully specified by a mean function and a covariance (or kernel) function that encodes assumptions about smoothness, correlations, and characteristic length scales of the target function \cite{Rasmussen2006}. We refer the reader to \ccite{Rasmussen2006} for a more in-depth introduction to GPR.

\subsubsection{\textbf{Kernel and mean function choices}}
GPRs can be constructed with many possible choices of kernels and mean functions. To capture both smooth, global trends and more localized structure, we adopt a mixed kernel formed by a weighted sum of a squared-exponential Radial Basis Function (RBF) kernel and a Matern kernel,

\begin{equation}
k(\mathbf{x},\mathbf{x}') = \alpha_1 k_{\text{RBF}}(\mathbf{x},\mathbf{x}') + \alpha_2 k_{\text{Matern}}(\mathbf{x},\mathbf{x}')
\end{equation}

Both kernels employ automatic relevance determination (ARD), allowing the model to learn independent characteristic length scales along each physical parameter dimension. The kernel weights $\alpha_1$, $\alpha_2$, together with all other kernel hyperparameters, are optimized by maximizing the marginal likelihood during training.

Rather than assuming a zero-mean prior, we use a linear mean function, which captures leading-order trends in the data and improves extrapolation behavior at the edges of the training domain. All input parameters and target quantities are standardized to zero mean and unit variance prior to training, with predictions subsequently transformed back to physical units. Given our dataset of only $O(10^3)$ simulations, the cubic scaling of exact GPR inference does not pose a computational limitation.

\subsubsection{\textbf{Implementation}}
Our GPR workflow is implemented in a \texttt{Python} package developed for this work and released as open source within the SXS \texttt{SimulationSupport} repository \cite{SimulationSupport}. The package provides the complete pipeline for data normalization, training, prediction, cross-validation, plotting, and analysis. Training and inference are performed using \texttt{PyTorch}~\cite{pytorch} and \texttt{GPyTorch}~\cite{gpytorch}, an open-source machine learning framework that supports automatic differentiation and GPU acceleration.

To make our model easy to use, we train it once and save it to disk. For each model, we store
\begin{enumerate}[label=(\arabic*)]
    \item the \texttt{PyTorch} \cite{pytorch} state dictionary of the trained GPR model, including the kernel hyperparameters and mean function parameters,
    \item the associated likelihood parameters,
    \item the input normalization statistics (the means and standard deviations for each feature),
    \item the output scaling used during training,
    \item the metadata.
\end{enumerate}
These quantities get written to a single file. A simple helper function reconstructs the full model architecture, restores all parameters, and returns objects ready for inference. Users can then simply supply their desired parameters and obtain GPR corrections in a single call, without needing to access the original training catalog or needing to train the model.

\subsubsection{\textbf{Model construction and training strategy}}
We train two independent GPR models: one to predict corrections to the initial orbital frequency, $\Omega_0$, and another to predict corrections to the radial expansion rate, $\dot{a}_0$. This choice allows each parameter to be modeled with its own characteristic scale and smoothness, and avoids introducing additional hyperparameters associated with multioutput kernels. Given the high predictive accuracy achieved for each parameter individually, we find no need to introduce cross-output covariance structure.

Each model takes as input a set of physical binary parameters, such as the mass ratio, $q$, initial separation, $\psep$, and the individual spin components, $(S_{1x}, S_{1y}, S_{1z}, S_{2x}, S_{2y}, S_{2z})$. Rather than predicting the orbital parameters directly, we instead build models for the difference between $\Omega_0$ and $\dot{a}_0$ and their PN values. This formulation focuses the regression on learning small, smooth corrections to the PN baseline rather than the full parameter values.

\subsubsection{\textbf{Validation and application}}
At each stage, we assess model performance using leave-one-out (LOO) cross validation, in which each simulation is withheld in turn from the training set and predicted by a model trained on the remaining data.

Our workflow proceeds as follows:
\begin{enumerate}[label=(\arabic*)]
\item For each simulation in the training set, we extract the intrinsic binary parameters, $q$, $\mathbf{S}_1$, $\mathbf{S}_2$, and the initial separation, $\psep$, together with the PN initial guesses for the orbital frequency, $\Omega_0^{\rm PN}$, and radial expansion rate, $\dot a_0^{\rm PN}$.
\item Using the final, low-eccentricity orbital parameters obtained after the \spec{} eccentricity-reduction procedure, we define the training data as residual corrections to the PN predictions,
\begin{align}
\Delta \Omega_0 &= \Omega_0^{\rm NR} - \Omega_0^{\rm PN}, \\
\Delta \dot a_0 &= \dot a_0^{\rm NR} - \dot a_0^{\rm PN}.
\end{align}
We always use the LOPN approximation for $\Omega_0^{\rm PN}$ and $\dot a_0^{\rm PN}$ in this work, but any smooth reference function could be used.
\item We train two independent GPR models to predict
$\Delta \Omega_0$ and $\Delta \dot a_0$ as smooth functions of the input parameters, assessing the strength of our model and its ability to generalize using LOO cross validation.
\item For a new target configuration, the trained GPR models predict corrections to the PN initial guesses, yielding corrected orbital parameters,
\begin{align}
\Omega_0 = \Omega_0^{\rm PN} + \Delta \Omega_0^{\rm GPR}, \\
\dot a_0 = \dot a_0^{\rm PN} + \Delta \dot a_0^{\rm GPR}.
\end{align}
These corrected parameters are used to construct NR initial data.
\end{enumerate}

\section{Results}
\label{sec:results}

\subsection{Equal-mass, nonspinning binaries}

We begin by constructing a deliberately simplified model restricted to the one-dimensional parameter space of equal-mass, nonspinning binaries, where the initial separation, $\psep$, is the sole input parameter, and the initial orbital frequency, $\Omega_0$, and radial velocity, $\dot{a}_0$, are the two model outputs. This test serves as an initial validation and pedagogical demonstration of the GPR framework. We then progressively incorporate systems with increasing mass ratio and spin complexity, expanding the input parameter space accordingly.

\begin{figure}[h]
\includegraphics[width=\columnwidth]{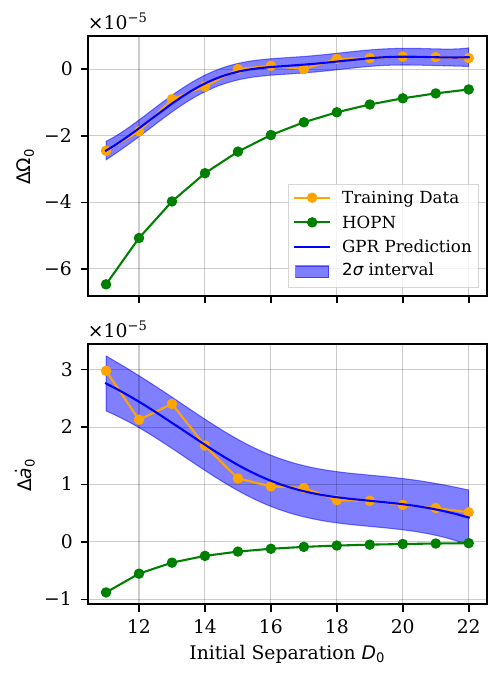}
\caption{\justifying GPR and HOPN predictions for low-eccentricity orbital parameters, $\Omega_0$ and $\dot{a}_0$, for equal-mass, nonspinning binaries. Plotted is the difference to the LOPN baseline function. The GPR prediction differs systematically from the HOPN prediction, particularly at smaller separations.}
\label{fig:equal-mass-nonspin}
\end{figure}

We perform eccentricity reduction on 12 equal-mass, nonspinning SKS initial data binary simulations with separations in the range of $11 \leq D/M \leq 22$, and find that despite the small training set, our model performs in agreement with the distribution across the parameter range. The corresponding predictions and confidence intervals are shown in \cref{fig:equal-mass-nonspin}, with the $\Delta$PN values shown by the 12 orange markers and orange curve, and the GPR predictive mean and associated uncertainty shown in blue.

The primary purpose of this test is not to achieve maximal predictive accuracy, but rather to verify that the GPR behaves sensibly in a controlled setting before extending the framework to higher-dimensional and more astrophysically relevant parameter spaces. In addition, to assess whether increasing PN order alone is sufficient for capturing the relevant structure, we also plot the prediction of the HOPN formulation across the same parameter space. As shown in \cref{fig:equal-mass-nonspin}, we find a systematic difference between the GPR predictions and the HOPN predictions, which we investigate in more detail in the next section.

\subsection{Unequal-mass, nonspinning binaries}

We next consider a more challenging test by extending the training data to a two-dimensional parameter space, introducing mass ratio, q, as an additional input parameter, alongside the initial coordinate separation, $\psep$. We continue to restrict to nonspinning configurations with SKS initial data, but now cover the ranges $12 \leq D/M \leq 22$ and mass ratios $1 \leq q \leq 8$, resulting in a dataset of 84 simulations.

As in the previous section, we take the LOPN prescription as the baseline and train the GPR model to learn the corrections to these PN initial guesses, and also plot the HOPN prediction. As shown in \cref{fig:unequal-mass-non-spinning}, the HOPN model predicts a nearly flat surface for both $\Omega_0$ and $\dot{a}_0$, indicating that the inclusion of higher-order terms leads to only minor modifications of the initial orbital parameters. In contrast, the GPR predictions exhibit a clear and systematic offset, with deviations that increase toward larger mass ratios and smaller separations. This qualitative difference demonstrates that the structure learned by the GPR is not captured fully by either PN formulation, even when higher-order terms are included. The discrepancy persists for both orbital parameters, indicating that the limitations of PN initial guesses in this regime cannot be resolved by simply increasing PN order. These results show that HOPN corrections and data-driven GPR corrections represent different modifications to the initial data. The GPR captures trends that are absent from analytic expansions, further motivating the use of data-driven approaches, even in parameter regimes where PN theory is already well behaved.

\subsection{Unequal-mass, aligned-spin binaries}

Having established that HOPN corrections fail to reproduce the structure learned by the GPR even in the nonspinning case, we now extend the analysis to spinning binaries, where analytic initial guesses are expected to be less reliable. As an intermediate step toward the fully generic configuration space, we restrict our attention to aligned-spin systems, which introduce additional physical complexity, while avoiding the complications associated with spin precession.

Specifically, we use data from the q83dAligned simulation campaign in the SXS catalog (SXS:BBH:1419--SXS:BBH:1509) \cite{sxscatalog}, consisting of nonprecessing binaries with mass ratios $1\leq q \leq 8$ and spins aligned with the orbital angular momentum. All simulations in this subset are generated using the same version of \spec{} and the SKS initial data prescription, ensuring that the numerical and gauge framework is consistent across the training set. We further choose the same high resolution for all simulations, denoted Lev3 in \spec{}. The resulting dataset consists of 90 unique, nonprecessing simulations with initial separations spanning $12 \leq D/M \leq 16$. Again, all simulations included in the training set have final eccentricities below $1\times10^{-3}$.

\begin{widetext}
\includegraphics[width=\columnwidth,clip,trim=0 2cm 0 4cm]{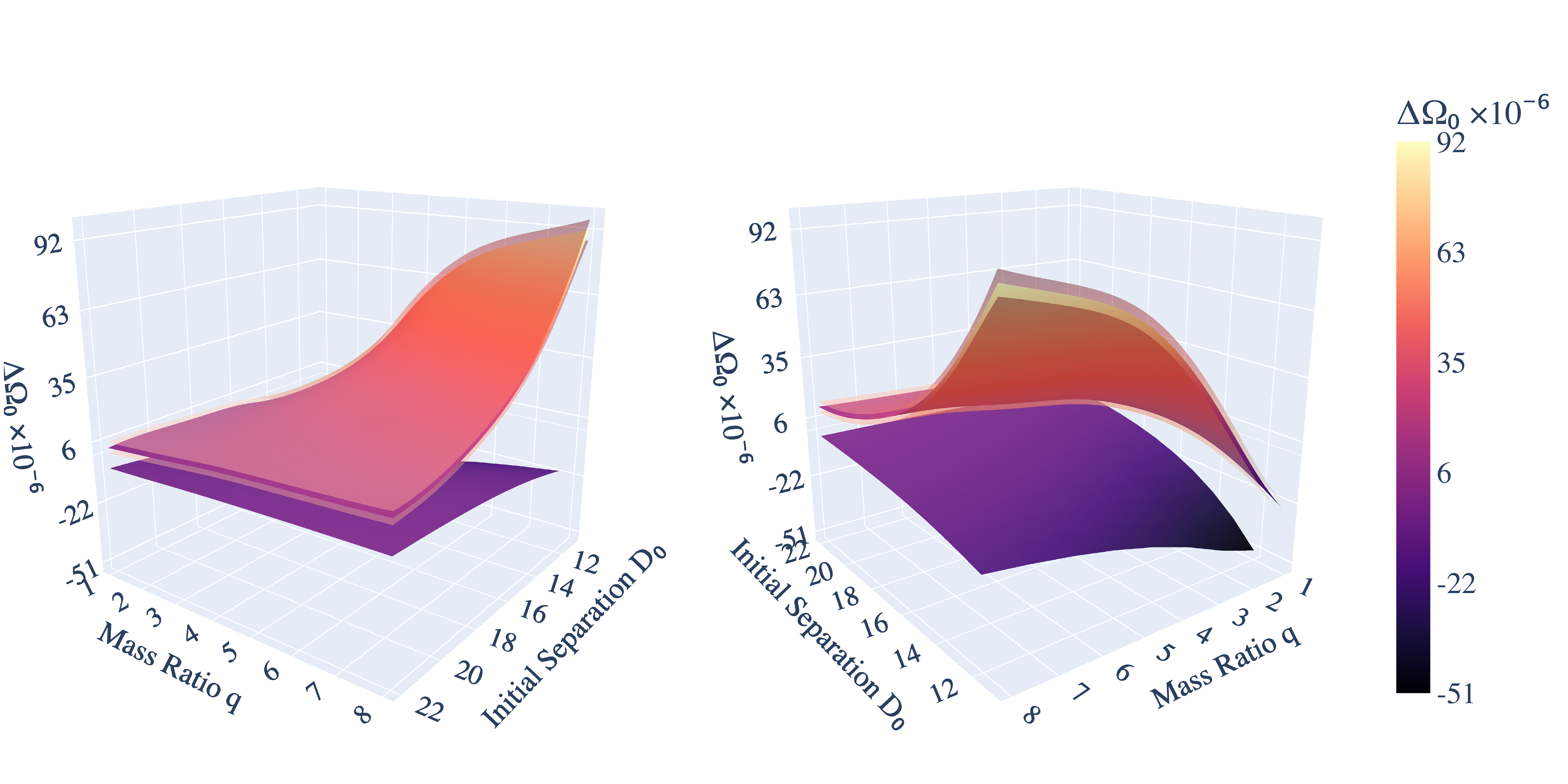} \\
\includegraphics[width=\columnwidth,clip,trim=0 2cm 0 4cm]{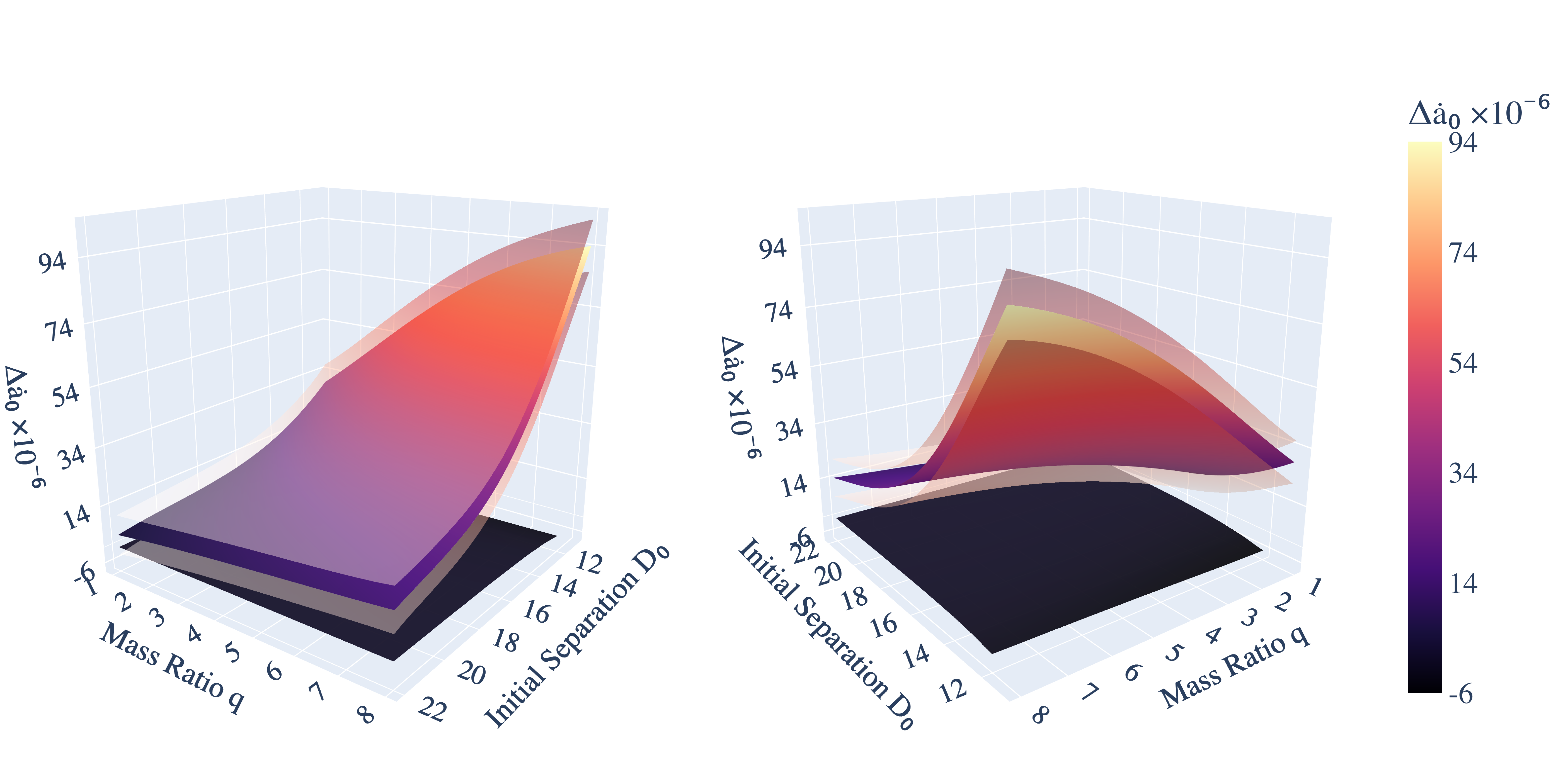}
\captionof{figure}{\justifying Comparison of GPR-predicted corrections and HOPN corrections for aligned-spin binaries as functions of initial separation, $\psep$, and mass ratio, $q$. Plotted is the difference to the LOPN baseline. The top panels show two views of the predictions for $\Omega_0$, while the bottom panels show two views of the predictions for $\dot{a}_0$. On all four plots, the upper surfaces show the GPR corrections, with the lighter shaded regions indicating $3\sigma$ confidence intervals, while the lower surfaces show the HOPN corrections. The GPR predictions differ systematically from the PN predictions.}
\label{fig:unequal-mass-non-spinning}
\end{widetext}

\begin{widetext}
    \includegraphics[width=\columnwidth]{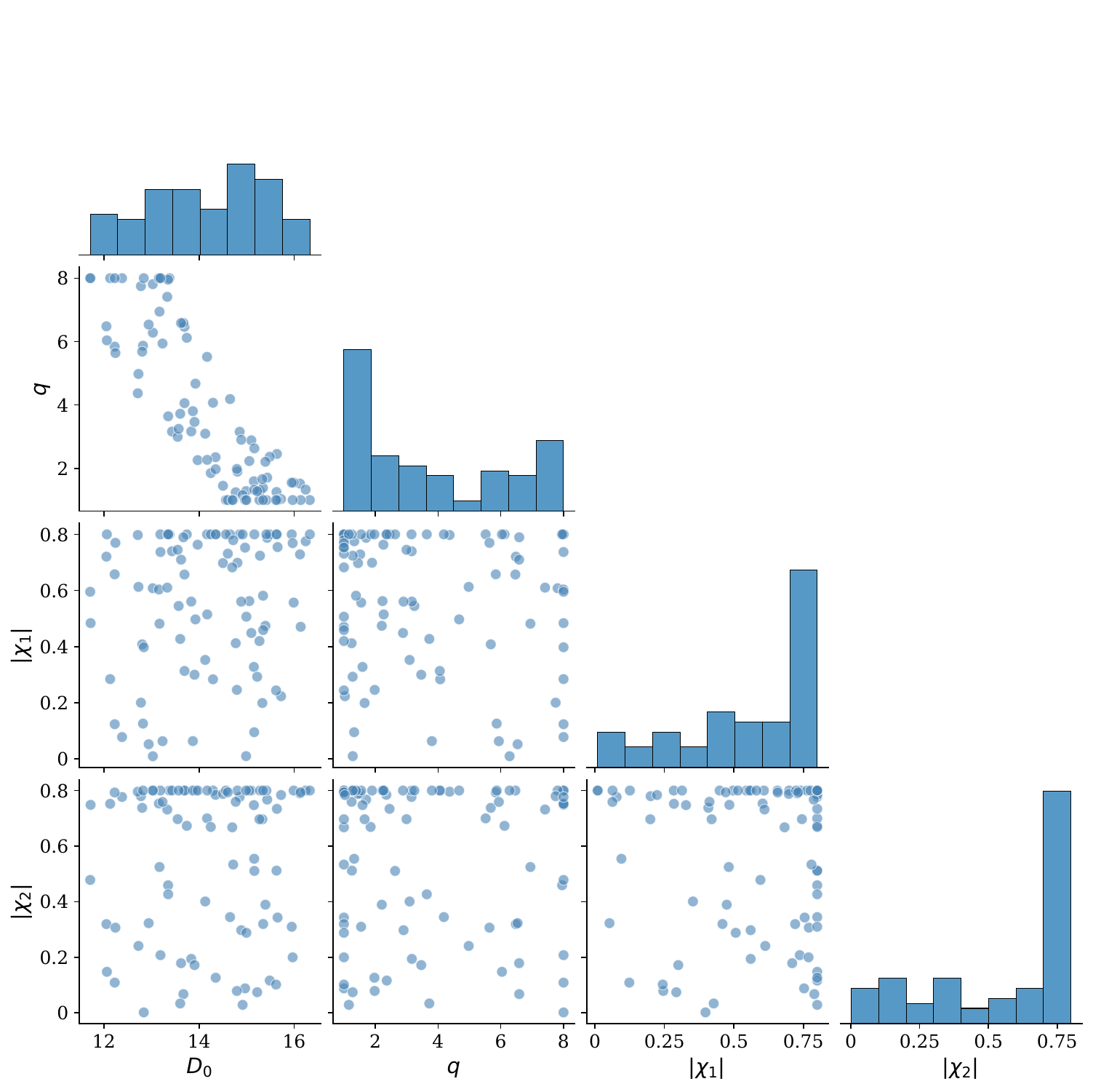}
    \captionof{figure}{\justifying Training dataset used for our four-dimensional GPR model. The parameters shown are the initial separation, $\psep$, mass ratio, $q$, and spin magnitudes, $|\chi_1|$ and $|\chi_2|$, for the 90 unique simulations taken from the q83dAligned subset of the SXS catalog, which had final eccentricities below $1\times10^{-3}$. The diagonal panels display one-dimensional histograms of each variable, while the off-diagonal panels show pairwise scatter plots illustrating the correlations among sets of parameters. These distributions define the region of parameter space on which our four-dimensional model is trained on.}
    \label{fig:q83d_params}
\end{widetext}

The coverage of the four-dimensional training set is summarized in \cref{fig:q83d_params}, which presents a corner plot of the joint and marginal distributions of the input parameters. The diagonal panels show the one-dimensional distributions of each parameter, while the off-diagonal panels illustrate pairwise correlations. Together, these distributions provide a compact visualization of the region of parameter space sampled by the training data and demonstrate that the dataset spans a broad and representative subset of the aligned-spin configuration space.

\begin{figure}[h]
    \includegraphics[width=\columnwidth]{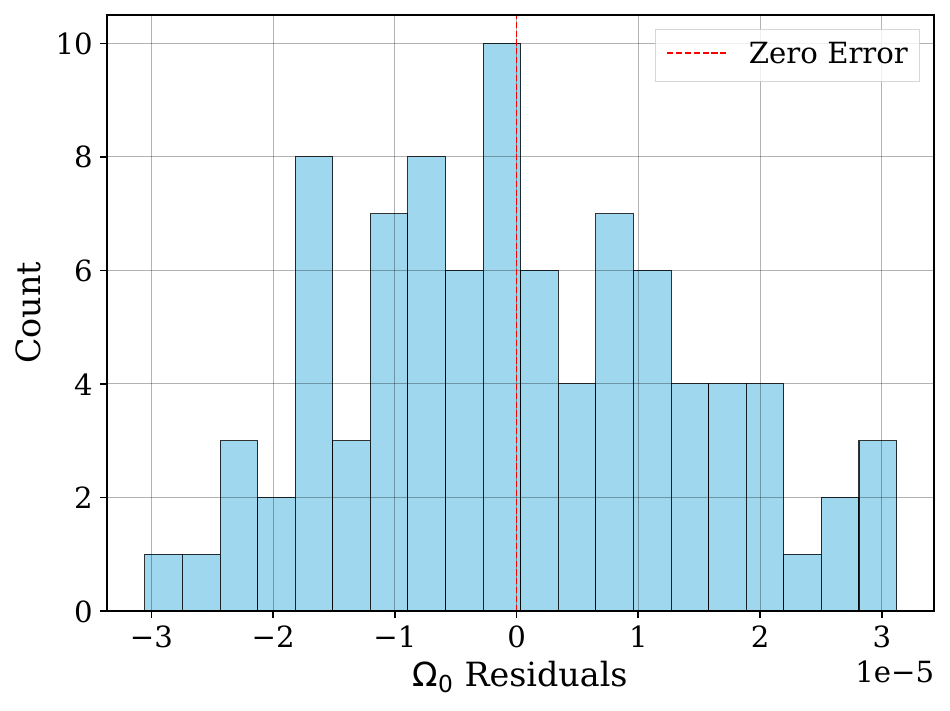} \\
    \includegraphics[width=\columnwidth]{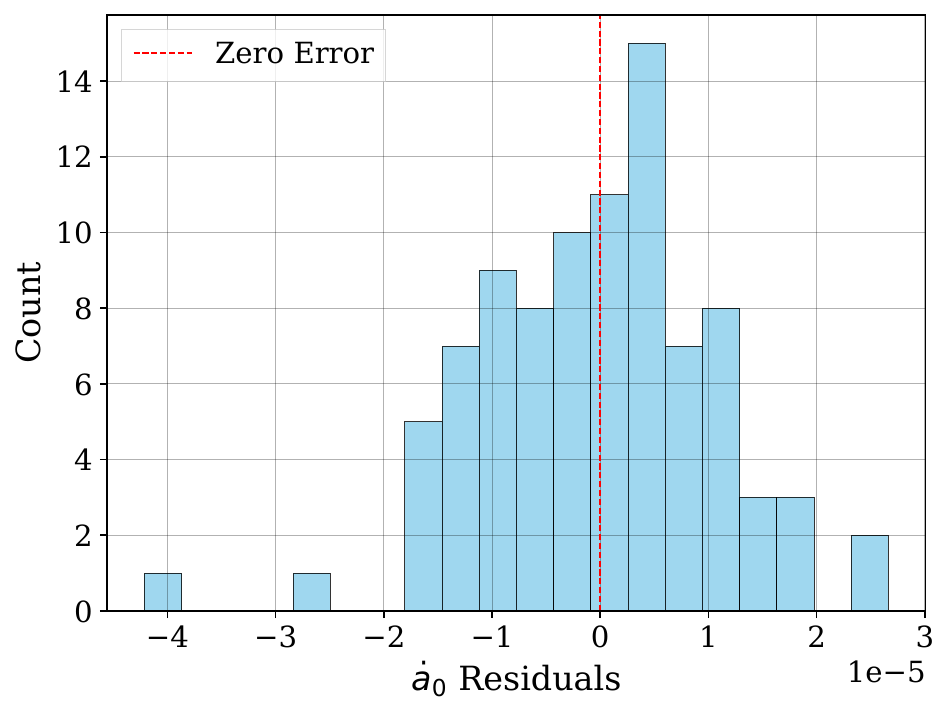}
    \caption{\justifying LOO residuals for the GPR-predicted corrections to $\Omega_0$ and $\dot{a}_0$, respectively, for the unequal-mass, aligned-spin binaries. Both histograms peak closely around the vertical dashed line, which marks zero error. We conclude that the GPR model is unbiased and does not systematically over- or under-predict the parameter corrections.}
    \label{fig:q83d_loo}
\end{figure}

As in the lower-dimensional cases, we train the GPR model on the residuals between the final simulation values and the LOPN baseline for $\Omega_0$ and $\dot{a}_0$. To assess the generalization performance of the model in this higher-dimensional setting, we perform a LOO cross validation across the full dataset. In this procedure, each simulation is withheld in turn, the model is retrained on the remaining 89 simulations, and the excluded point is used to evaluate the prediction error. This approach provides an unbiased estimate of predictive performance across the sampled parameter space.

The resulting LOO residual distributions for both $\Omega_0$ and $\dot{a}_0$ are shown in \cref{fig:q83d_loo}. In both cases, the residuals are narrowly distributed and centered around zero, indicating that the GPR predictions are unbiased. The spread of the residuals is small compared to the overall range of the target values, with root-mean-square and mean absolute errors well below $1\%$ of the target range, and coefficients of determination exceeding $R^2=0.99$. The full numerical values for the cross-validation metrics are provided in Table \ref{tab:LOO_4d} of \cref{app:LOO_4d}. The absence of significant outliers suggests that the model generalizes smoothly across the sampled space, rather than relying on localized interpolation around individual training points. The symmetric, bell-shaped curve follows a Gaussian.

This aligned-spin, four-dimensional test serves as a critical validation of the stability and scalability of our framework. Having demonstrated robust performance in this intermediate setting, we are now able to extend the model to the full, eight-dimensional parameter space of generic spinning binaries.

\subsection{Unequal-mass, precessing-spin binaries}
We next extend the GPR framework to the full, eight-dimensional parameter space by incorporating the complete three-dimensional spin vectors of both black holes, in addition to the initial separation and mass ratios. This model therefore captures the full spin dependence of the binary, including spin orientations and precession effects.

We utilize the q87d simulation campaign of the SXS catalog that was performed in 2019 (SXS:BBH:263--SXS:BBH:3617) \cite{sxscatalog}, consisting of precessing binaries with mass ratios $1\leq q \leq8$ and generic spin configurations, again generated using the same version of \spec{} and the SKS initial-data prescription. We again choose the same Lev3 high resolution for all simulations, resulting in a dataset of 958 unique simulations with final eccentricities below $1 \times 10^{-3}$.

\begin{widetext}
    \includegraphics[width=\textwidth]{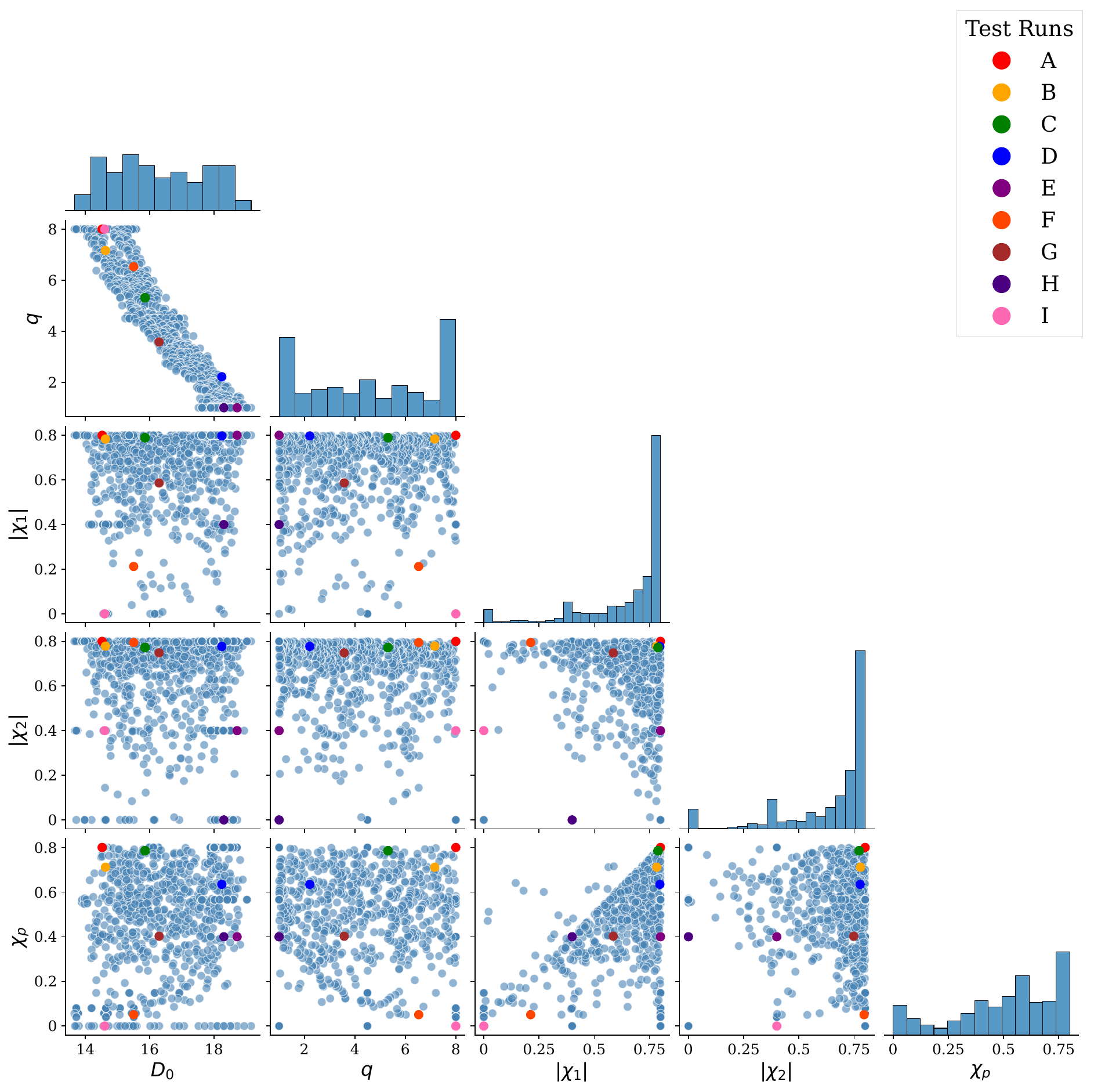}
    \captionof{figure}{Training dataset used for our 8D GPR model. Shown are initial separation, $\psep$, mass ratio, $q$, spin magnitudes, $|\chi_1|$ and $|\chi_2|$, and the effective precession spin, $\chi_p$, for the 958 unique simulations. The diagonal panels show one-dimensional histograms of each variable, while the off-diagonal panels show pairwise scatter plots illustrating the correlations between sets of parameters. The colored markers indicate specific test runs utilized later for further model evaluation. The training set spans a broad range of physical configurations, and the selected test cases probe different regions of that space.}
    \label{fig:q87d_cornerplot}
\end{widetext}

The coverage of parameter space is summarized in \cref{fig:q87d_cornerplot}, which presents a corner plot of the joint and marginal distributions of the input parameters. The diagonal panels show the one-dimensional distributions of the initial separation, mass ratio, spin magnitudes, and effective precession, while the off-diagonal panels show the corresponding pairwise correlations. The highlighted points represent data utilized later as test runs to validate our model, discussed in detail in the next section. As in the lower-dimensional cases, we train the GPR models on the residuals between the final simulation values and the LOPN baseline, extending our code to now handle eight input parameters: $\psep$, $q$, and the three spin components of each black hole, while retaining the two orbital parameters as outputs.

\begin{figure}
    \subfloat{\includegraphics[width=\columnwidth]{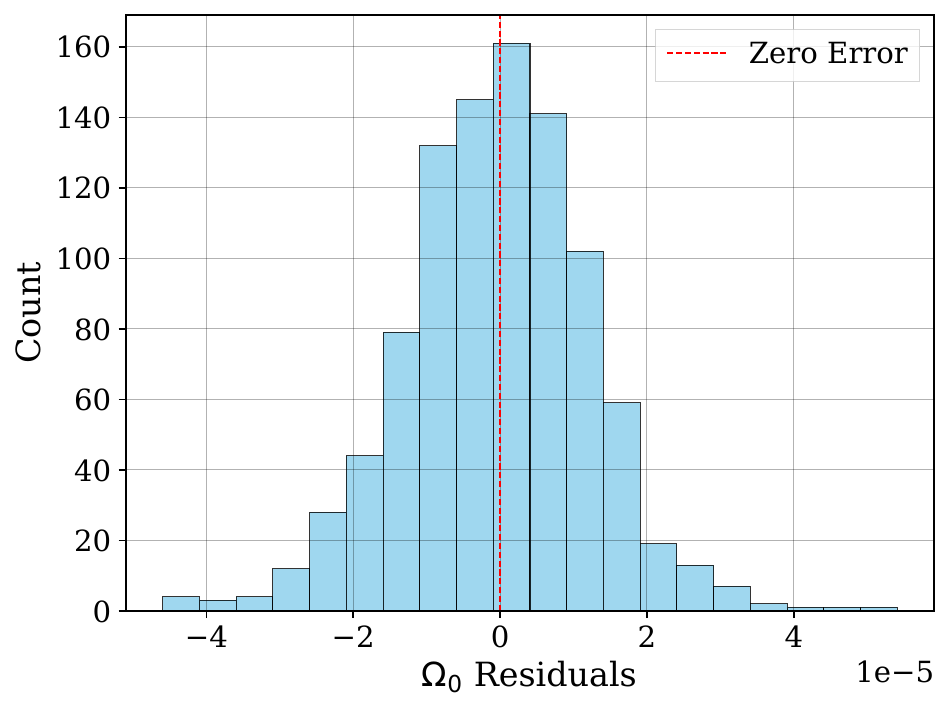}} \\
    \subfloat{\includegraphics[width=\columnwidth]{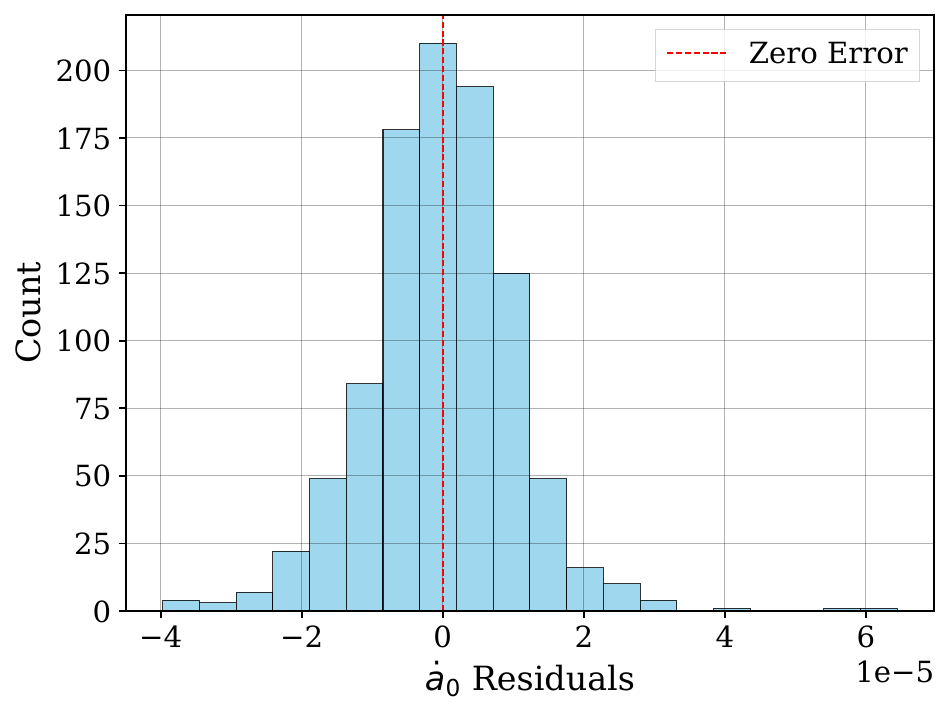}}
    \caption{\justifying Leave-one-out residuals for the GPR-predicted corrections to $\Omega_0$, and $\dot{a}_0$, respectively, for the eight-dimensional model. In both cases, the residuals are narrowly distributed and centered around the dashed vertical lines, which mark zero error, indicating that the GPR models are effectively unbiased. The symmetric, Gaussian shapes and absence of large outliers demonstrate stable performance and generalization across the full parameter space. The narrow width of the distributions relative to the physical scale of the correction indicates high predictive accuracy.}
    \label{fig:q87d_loo}
\end{figure}

Despite the increased dimensionality and inclusion of precessional dynamics, the GPR maintains stable and unbiased performance across the dataset. To validate these results, we again perform a leave-one-out experiment, training the model on 957 points and evaluating on the remaining one, repeating the process for all 958 simulations in the dataset. The resulting LOO predictions yield residuals centered around zero with no evidence of systematic bias, with coefficients of determination $R^2 > 0.99$ for both orbital parameters. Figure~\ref{fig:q87d_loo} shows histograms of the LOO residuals for $\Omega_0$ and $\dot{a}_0$, respectively. The full details of the LOO metrics are shown in \cref{tab:LOO_8d} of \cref{app:LOO_8d}. These results indicate that the data-driven approach remains stable and predictive even in the fully generic eight-dimensional parameter space.

\subsection{Validation on new simulations}

Above we compared the GPR predictions with the results of previously carried out eccentricity reductions for the same initial guesses. Here, we further validate the GPR framework by repeating the eccentricity-reduction procedure on a sample of configurations that we individually remove from the training set. These validation runs differ from the training data in two important ways. First, we use the current (2025) version of the eccentricity-reduction procedure, which incorporates several updates relative to earlier implementations (see Ref.~\cite{Habib2025}). Second, for each configuration we perform two independent eccentricity-reduction sequences under identical conditions: one initialized from the PN initial guess, and one initialized from the GPR-predicted parameters. The test runs are selected to span a range of configurations within the parameter space. Their locations are highlighted in the previously discussed corner plot of \cref{fig:q87d_cornerplot} and are labeled A--I.

\begin{figure}
    \includegraphics[width=\columnwidth]{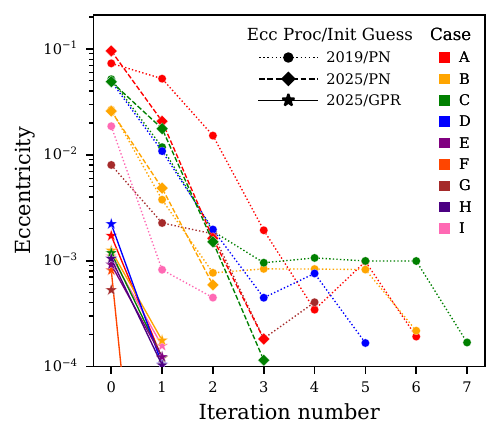}
    \caption{\justifying Eccentricity as a function of iteration number for nine test cases, labeled A--I. Dotted lines with circles correspond to PN initialized simulations run with the 2019 eccentricity-reduction procedure, dashed lines with diamonds correspond to PN initialized reruns of these simulations in 2025 with an updated eccentricity-reduction procedure (see Ref. \cite{Habib2025}), and solid lines with stars correspond to GPR initialized simulations run with the 2025 eccentricity-reduction procedure, trained on the 2019 data. In all nine cases, the GPR initialized runs reach the target eccentricity of $1\times10^{-3}$ either immediately or within a single iteration, while the PN initialized runs require multiple iterations. The initialization details of these runs are shown in \cref{tab:sim_info}.}
    \label{fig:q87d_iterations}
\end{figure}

Figure~\ref{fig:q87d_iterations} compares the number of eccentricity-reduction iterations required to reach a target eccentricity of $1\times10^{-3}$ for these simulations.
In all cases, the PN initialized runs require multiple iterations to converge, up to seven with the 2019 eccentricity-reduction procedure, and up to three with the updated 2025 procedure. In contrast, the GPR-based initial guesses achieve the target eccentricity either immediately, or within a single iteration. Details of these cases are shown further in \cref{tab:sim_info}. This consistent improvement across all tested configurations, despite variations in separations, mass ratios, and spin, demonstrates that the GPR model does effectively generalize to new simulations and provides a substantial improvement over existing PN initial guesses.

Figure~\ref{fig:trajectories} illustrates this behavior with a comparison of the trajectories of three of these test cases in the ($\Omega_0, \dot{a}_0$) plane. Each numbered point corresponds to one iteration of the eccentricity-reduction procedure, showing the successive updates to the orbital parameters. The insets zoom in on the final iterations, where the trajectories converge. 

To quantify the accuracy of the final iteration, we define a tolerance region around the converged point. This region is constructed from the difference between the final orbital parameters and the next updated and unused parameters that would have been applied had the eccentricity threshold not already been satisfied. In practice, this provides an estimate of the local sensitivity of the eccentricity-reduction procedure near convergence. This estimate provides only a lower bound of the tolerance region, with the true tolerance region expected to be strictly larger, assuming the eccentricity-reduction procedure is still converging. Therefore, we plot the estimated tolerance regions as ellipses centered on the final point for the 2025 runs, with sizes corresponding to $1\times$, $2\times$, and $3\times$ this step size.

While the new (2025) PN initialized runs converge faster than their 2019 counterparts, the initial guesses remain essentially the same, highlighting an important point: in the past six years, there have been many improvements in \spec{} and in the eccentricity fitting algorithm now implemented in \spec{} (see Ref.\cite{Habib2025}), but adding higher-order terms did not significantly alter the initial guess itself. In contrast, the GPR predictions place the initial parameters close to the final, low-eccentricity points of the training data, and reach the eccentricity threshold in one iteration. Additionally, the endpoints of both the 2025 PN and the GPR initialized runs lie within, or near each other's estimated tolerance region, whereas the six-year-old training data are consistently slightly different.

Despite being trained on simulations generated six years earlier, the GPR model consistently provides accurate predictions, and significantly reduces the need for iterative eccentricity reduction. \cref{tab:sim_info} displays the total run-time and total number of iterations for the nine test cases, further highlighting that the GPR initialized simulations are consistently faster than their PN counterparts. All of our simulations were performed on a single node with 56 Intel Skylake CPU cores in the Resnick High Performance Computing Center at Caltech.

\subsection{Extrapolation performance in mass ratio}
To assess the extrapolation capabilities of the GPR model beyond its training domain, we conduct a case study in mass ratio. Our training data again span mass ratios $1\leq q \leq 8$. For each chosen cutoff in the range $q_{\mathrm{cutoff}} \in \{2 , 2.5 ..., 8\}$ with steps of size $0.5$, we train a GPR model using only simulations with $q \leq q_{\mathrm{cutoff}}$ and test it on the held-out data, $q> q_{\mathrm{cutoff}}$.

Many simulations in the catalog \cite{sxscatalog} were intended to have integer or half-integer values of $q$, but since $q$ is measured at a reference point early in the simulation, as opposed to being exactly specified, the actual values of $q$ differ from their intended values by a small fraction on the order of $10^{-4}$ or smaller \cite{sxscatalog}. To account for this, we widen the cutoff slightly and include simulations with $q \leq q_{\mathrm{cutoff}}+ \Delta q$,
where $\Delta q = 10^{-2}$.

For each choice of $q_{\mathrm{cutoff}}$, we compute the worst-case prediction error across all test points using the absolute max norm,
\begin{equation}
\begin{aligned}
\label{eq:maxNorm}
\| \Delta \Omega_0 \|_{\infty} = \max_i |\Omega_{\mathrm{pred},i} - \Omega_{\mathrm{ref},i}| \\
\| \Delta \dot{a}_0 \|_{\infty} = \max_i |\dot{a}_{\mathrm{pred},i} - \dot{a}_{\mathrm{ref},i}|,
\end{aligned}
\end{equation}
which allows us to quantify the single largest deviation between the GPR prediction and the chosen reference. When comparing against the full-range GPR trained on the complete dataset, we take $\Omega_{\mathrm{ref}} = \Omega_{\mathrm{full\ GPR}}$ and $\dot{a}_{\mathrm{ref}} = \dot{a}_{\mathrm{full\ GPR}}$, and take the norm over all training points.

\begin{widetext}
    \includegraphics[width=\columnwidth]{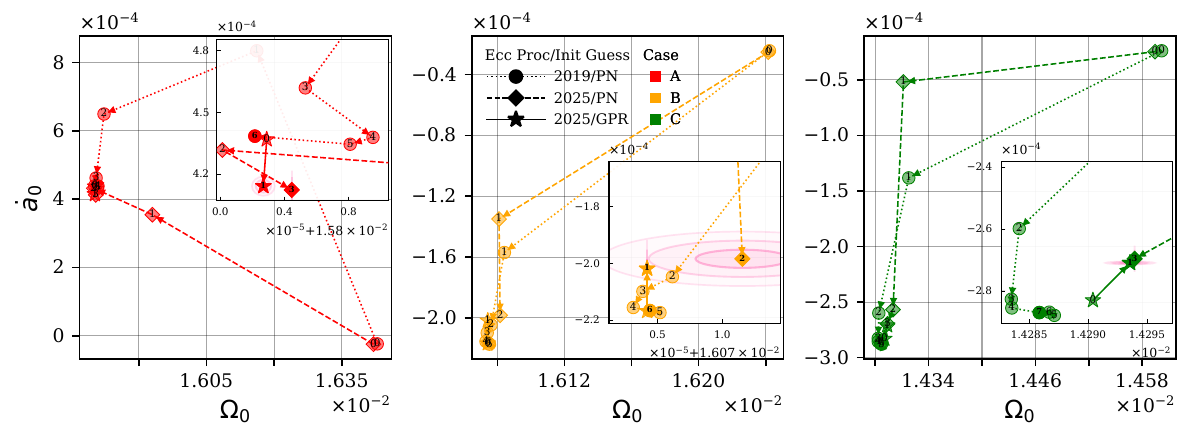}
    \captionof{figure}{\justifying Trajectories of the eccentricity-reduction procedure in the $(\Omega_0, \dot{a}_0)$ plane for three test cases (A--C). Each numbered point corresponds to one iteration of eccentricity reduction. The dotted lines with circular markers denote PN initialized simulations carried out with the 2019 eccentricity-reduction procedure, dashed lines with diamond markers denote PN initialized reruns of these simulations carried out with the 2025 eccentricity-reduction procedure~\cite{Habib2025}, and solid lines with star markers denote GPR initialized simulations carried out with the 2025 eccentricity-reduction procedure. The insets enlarge the final iterations, with estimated tolerance regions of sizes 1x, 2x, and 3x plotted around the final point. The GPR model predicts parameters near the training data (end points of dotted lines/circles) and then takes a single step to reduce the eccentricity below the threshold (solid lines/stars). The end points of the two 2025 runs (GPR initialized and PN initialized) lie within or near each other's estimated tolerance regions, whereas the 2019 training data is consistently slightly different due to the mentioned changes in \spec{} since 2019.}
    \label{fig:trajectories}

    \captionof{table}{Test cases A--I, identified by their corresponding SXS IDs, mass ratios (q), and spin magnitudes ($|\chi_1|$, $|\chi_2|$). For each case, ``Initialization'' describes the method utilized to generate the initial guess (PN or GPR), and ``Ecc Procedure'' gives the year the simulation was run (2019 or 2025). Columns ``Ecc0''--``Ecc3'' give the wall-clock time (rounded to the nearest hour) for each successive eccentricity-reduction iteration. The column ``Total Time'' is the sum of the ``Ecc0''--``Ecc3'' columns (computed from the unrounded values, then rounded to the nearest hour). The column ``Total Iterations'' counts the total number of eccentricity-reduction iterations (excluding the initial, required Ecc0).}
    \setlength{\tabcolsep}{2pt}
    \begin{tabular}{ccccccccccccc}
    \toprule
    Case & SXS ID & $q$ & $|\chi_1|$ & $|\chi_2|$ & Initialization & Ecc Procedure & Ecc0 & Ecc1 & Ecc2 & Ecc3 & Total time & Total Iterations \\
    \midrule
    \multirow{3}{*}{A} & \multirow{3}{*}{SXS:BBH:2728} & 
    \multirow{3}{*}{8.00} & \multirow{3}{*}{0.80} & \multirow{3}{*}{0.80} & PN & 2019 & - & - & - & - & - & 6 \\
    & & & & & PN & 2025 & 321 & 309 & 321 & 311 & 1260 & 3 \\
    & & & & &  GPR & 2025 & 314 & 313 & - & - & 628 & 1 \\ 
    \midrule
    \multirow{3}{*}{B} & \multirow{3}{*}{SXS:BBH:2888} & \multirow{3}{*}{7.16} & \multirow{3}{*}{0.78} & \multirow{3}{*}{0.78} & PN & 2019 & - & - & - & - & - & 6 \\
    & & & & & PN & 2025 & 218 & 224 & 230 & - & 672 & 2 \\
    & & & & & GPR & 2025 & 232 & 232 & - & - & 464 & 1 \\
    \midrule
    \multirow{3}{*}{C} & \multirow{3}{*}{SXS:BBH:3066} & \multirow{3}{*}{5.31} & \multirow{3}{*}{0.79} & \multirow{3}{*}{0.77} & PN & 2019 & - & - & - & - & - & 7 \\
    & & & & & PN & 2025 & 161 & 163 & 171 & 166 & 661 & 3 \\
    & & & & & GPR & 2025 & 168 & 167 & - & - & 335 & 1 \\
    \midrule
    \multirow{2}{*}{D} & \multirow{2}{*}{SXS:BBH:3380} & \multirow{2}{*}{2.21} & \multirow{2}{*}{0.80} & \multirow{2}{*}{0.78} & PN & 2019 & - & - & - & - & - & 5 \\
    & & & & & GPR & 2025 & 82 & 81 & - & - & 163 & 1 \\
    \midrule
    E & SXS:BBH:3604 & 1.00 & 0.80 & 0.40 & GPR & 2025 & 35 & 35 & - & - & 70 & 1 \\
    \midrule
    F & SXS:BBH:2931 & 6.53 & 0.21 & 0.80 & GPR & 2025 & 214 & 210 & - & - & 424 & 1 \\
    \midrule
    G & SXS:BBH:3236 & 3.58 & 0.59 & 0.75 & GPR & 2025 & 95 & - & - & - & 95 & 0 \\
    \midrule
    H & SXS:BBH:3616 & 1.00 & 0.40 & 0
    & GPR & 2025 & 16 & 16 & - & - & 32 & 1 \\
    \midrule
    \multirow{2}{*}{I} & \multirow{2}{*}{SXS:BBH:2755} & \multirow{2}{*}{8.00} & \multirow{2}{*}{0} & \multirow{2}{*}{0.40} & PN & 2019 & - & - & - & - & - & 2    \\
    & & & & & GPR & 2025 & 90 & 91 & - & - & 181 & 1 \\
    \bottomrule
    \end{tabular}
    \label{tab:sim_info}
\end{widetext}

\begin{figure}
    \includegraphics[width=\columnwidth]{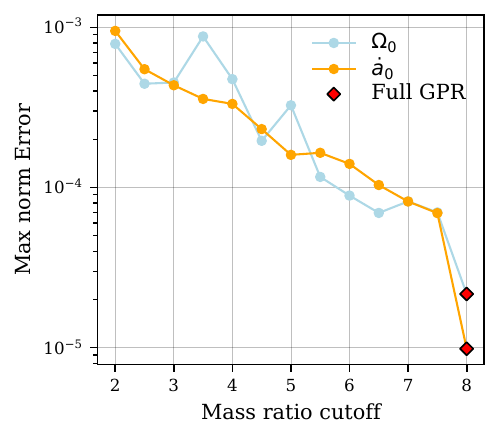}
    \caption{\justifying Mass ratio extrapolation tests of the GPR model. Each point represents the worst-case deviation for that cutoff, and the blue and orange curves correspond to $\Omega_0$ and $\dot{a}_0$, respectively. We increment each testing cutoff by 0.5, until we reach $q \leq 8$ and are training on all 958 simulations. The red diamond represents the reference, full-range, nonextrapolated GPR error for mass ratio 8. We see an approximately exponential decrease in error with extrapolation distance in mass ratio.}
    \label{fig:extrapolation}
\end{figure}

Figure~\ref{fig:extrapolation} displays this mass-ratio extrapolation test of the GPR model. The first cutoff is $q \leq 2$ (plus the added tolerance of $10^{-2}$), providing 221 training simulations and 737 test simulations. The second cutoff is $q \leq 2.5$, providing 263 training simulations and 695 test simulations. We continue incrementing each cutoff by 0.5 until we reach $q \leq 8$ and are training on the full 958 simulations. The red diamond at $q=8$ is the nonextrapolated, full-range GPR error prediction. Although optimal performance is achieved when training on the full range of mass ratios, for both orbital parameters the errors decrease exponentially with extrapolation distance.

\begin{figure}
    \includegraphics[width=\columnwidth]{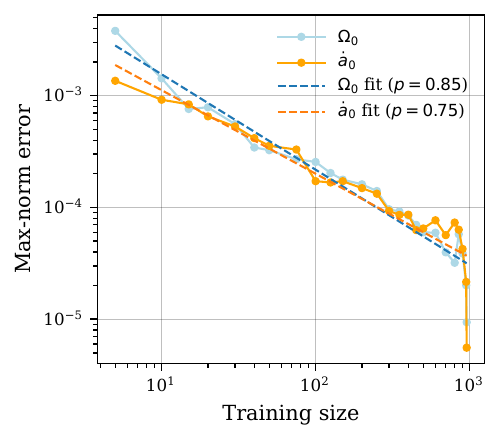}
    \caption{\justifying Convergence of our GPR model with training size. Plotted is the maximum norm error of the GPR-predicted $\pomega$ and $\padot$, blue and yellow, respectively. We show the power law convergence fits for both parameters. Both targets exhibit a nearly-linear decay in error as training size, $N$, increases, with estimated convergence rates between $\mathord{\sim}0.75$ and 0.85.}
    \label{fig:batches}
\end{figure}

\subsection{Dependence on training size}
We also conduct a series of tests to quantify how much data are required for accurate predictions and to identify where performance begins to saturate. For this test, we select a batch of $N_{\mathrm{train}}$ random training points from our full set of 958 simulations, train the GPR on this dataset, and evaluate the predictions on the remaining held-out runs. The procedure is repeated five times for robustness, and the absolute maximum norms taken over the test points are calculated for each batch. As shown in \cref{fig:batches}, we see an approximately linear decrease in error with training size.

At small training sizes, the largest errors occur at the edges of the mass ratio domain, where $q=8$, while at larger training sizes, the errors shift toward equal-mass cases of $q=1$, which are densely sampled. This transition indicates that the remaining error is not dominated by sparse coverage in $q$, but rather by intrinsic model variance instead of extrapolation.

Quantitatively, the worst-case deviation defined in \cref{eq:maxNorm} exhibits a power-law decay with training set size, $\epsilon \mathord{\sim} N^{-\alpha}$, with fitted exponents $\alpha = 0.85$ for $\Omega_0$ and $\alpha=0.75$ for $\dot{a}_0$. Such power-law learning curves are common for Gaussian process regression in finite-data regimes {\cite{Williams2000, Sollich2002, LeGratiet2015}} and are consistent with the smoothness of the target functions and use of ARD kernels, which effectively reduce the dimensionality of the problem {\cite{Rasmussen2006}}. This behavior is empirical and specific to the parameter range explored here.

\section{Discussion}
\label{sec:discussion}
The results presented above demonstrate that a GPR can provide an efficient framework for accurately predicting the orbital parameters required to obtain low-eccentricity initial data. While this method substantially reduces the cost of traditional eccentricity-reduction procedures, it also opens the door for several avenues for further developments. In this section, we discuss the potential uses of the framework, as well as limitations and directions for future work. 

\subsection{Large simulation campaigns}
One of the primary applications of the GPR framework developed in this work is its integration into large-scale NR simulation campaigns. Rather than requiring a fully trained model \textit{a priori}, the GPR can be constructed iteratively as simulations are generated. An initial batch of simulations located at the edges of the targeted parameter space can be run using the full eccentricity-reduction procedure, perhaps initialized with extrapolation from an existing GPR or with PN predictions. The resulting low-eccentricity orbital parameters then serve as training data for the new GPR. Subsequent batches of simulations can leverage the trained GPR to obtain significantly improved initial guesses for $\Omega_0$ and $\dot{a}_0$, reducing or fully eliminating the need for multiple eccentricity-reduction iterations. As additional simulations are completed, their results can be incorporated into the training set, progressively refining the model. In this way, the computational cost invested in early simulations is amortized over the entire campaign, enabling faster convergence for later runs. 

A closely related application is the use of the GPR to densely populate regions of parameter space that have already been explored. This is useful for producing higher-accuracy simulations with similar parameters, extending existing waveform catalogs, or improving the accuracy of surrogate models, where many simulations are required in a localized region of parameter space.

\subsection{Dependence on initial data choices}
An interesting next question is to what extent a GPR model trained on one class of initial data would be able to generalize to others. In this work, all training data is drawn from simulations constructed using a consistent initial data prescription (SKS initial data). In practice, NR simulations may employ different choices, such as a superposed harmonic Kerr conformal background \cite{Varma2018, Ma2021} or negative-expansion boundary conditions \cite{Varma2018}. The choice of initial data can lead to systematic differences in the resulting orbital parameters \cite{Pfeiffer2002}. It is not yet clear whether a model trained on one class of initial data can accurately predict corrections for another, or whether separate models are required. While the smooth dependence of the corrections on physical parameters suggests some degree of robustness, this assumption remains to be tested. This systematic study of cross-generalization between different initial data formulations would be interesting future work.

\subsection{Extension to additional quantities}
Although the present study focuses on predicting corrections only to the initial orbital frequency and radial velocity, the GPR framework is not inherently limited to these quantities. We now have a data-driven framework that can, in principle, be utilized to predict other simulation-level properties, such as the number of inspiral orbits or the time to merger. While we leave this investigation as future work, these could also be interesting properties to study.

\subsection{Limitations due to data quality}
It is important to note that the accuracy achievable by the GPR model is fundamentally limited by the quality of the training data. In particular, the model can only reproduce corrections consistent with the eccentricity tolerance achieved in the underlying simulations. For the purposes of our model, eccentricity $ < 1 \times 10^{-3}$ is treated as zero eccentricity. Further improvements in predictive accuracy will, therefore, eventually be limited by eccentricity tolerances in the training data, and it will likely become necessary to advance eccentricity-reduction techniques and/or reduce junk radiation present in the initial data. While these issues are of significant importance--especially for applications such as \textit{LISA} waveform modeling and PN-NR hybridization--they are beyond the scope of the present work.

These considerations illustrate how data-driven eccentricity control can be incorporated into existing NR workflows, while also highlighting the assumptions and limitations that must be addressed in future studies.

\section{Conclusion}
\label{sec:conclusion}
We have presented a data-driven approach for accelerating eccentricity reduction in numerical relativity simulations of binary black holes. By training a Gaussian process regression model on residuals between a post-Newtonian baseline and the final, low-eccentricity orbital parameters obtained from preexisting simulations, we can predict near-optimal parameters directly from the physical parameters of the binary.

In the process of building our GPR model, we also investigated whether utilizing a higher-order post-Newtonian formulation would be sufficient for predicting the corrections to the orbital parameters, instead of a GPR. We find that extending the PN expansion to higher order produces only minor changes in the predicted values of $\Omega_0$ and $\dot{a}_0$. In contrast, the GPR model captures  systematic differences between PN predictions and the numerical relativity simulations, including effects arising from the initial-data construction and early gauge dynamics, that are not fully captured by PN approximations alone.

We validated our approach across the full parameter space of unequal masses up to mass ratio 8, initial separations between 11 and 22M, and precessing spins. Our cross validation and tests on withheld simulations further confirm that the model generalizes robustly across the full parameter space. The framework is able to consistently reduce the number of eccentricity-reduction iterations to zero or one, yielding a substantial reduction in computational cost, relative to standard post-Newtonian initializations \cite{Buonanno2011, Buchman2012}. Each additional eccentricity-reduction iteration increases total computational cost by roughly $10\%$, so a simulation requiring five eccentricity-reduction iterations incurs a total computational cost of $1.5\times$ that of an identical simulation initialized at sufficiently low eccentricity.

Even with six-year-old data, this approach eliminates the need for repeated simulation cycles, while maintaining accuracy. Our method does not modify the definition or measurement of eccentricity, nor does it replace existing eccentricity-reduction procedures. Instead, it accelerates an already established pipeline and inherits the diagnostic and gauge dependencies developed in years of previous work \cite{Buonanno2011, Pfeiffer2007, Islam2025, Knee2022, Habib2025, Zhang2013, HusaRamos2008, Ramos2019}.

While our study focuses on simulations produced using the \spec{} framework and SXS catalog \cite{spec, sxscatalog}, the underlying strategy is general and has the potential to be extended to other numerical-relativity formulations, such as \spectre{} \cite{spectre}. More broadly, we illustrate the power of machine learning techniques to accelerate and streamline computationally expensive components of numerical-relativity workflows without altering their physical assumptions. This supports ongoing efforts to expand waveform catalogs for current and future gravitational-wave observatories \cite{sxscatalog, Porter2010, Key2011, ligo, virgo}.

\begin{acknowledgments}
This work is supported in part by NSF Grants No. PHY-2309211, No. PHY-2309231, and No. OAC-2513339 and NASA Award No. 80NSSC26K0340 at Caltech; by NSF Grants No. PHY-2308615 and No. OAC-2513338 and NASA Award No. 80NSSC26K0340 at Cornell; and by the Sherman Fairchild Foundation at Caltech and Cornell. N.L.V. acknowledges support from the Swiss National Science Foundation (SNSF) Ambizione Grant No. PZ00-2\_232961.
\end{acknowledgments}

\section*{DATA AVAILABILITY}
The data that support the findings of this article are openly available \cite{SimulationSupport}.

\appendix
\begin{widetext}
\section{LEAVE-ONE-OUT CROSS-VALIDATION METRICS FOR THE FOUR-DIMENSIONAL GPR MODELS}
\label{app:LOO_4d}
    \captionof{table}{Shown are the root-mean-square error (RMSE), mean absolute error (MAE), coefficient of determination ($R^2$), mean, standard deviation (SD), maximum, and minimum.}
    \label{tab:LOO_4d}
    \setlength{\tabcolsep}{8.5pt}
    \begin{tabular}{cccccccc}
    \hline Quantity & RMSE & MAE & $R^2$ & Mean & SD & Max & Min\\
    \hline $\Delta\Omega_0$ & $1.4\times10^{-5}$ & $1.2\times10^{-5}$ & 0.996 & $2.65\times10^{-7}$ & $1.41\times10^{-5}$ & $3.12\times10^{-5}$ & $-3.05\times10^{-5}$ \\ $\Delta\dot{a}_0$ & $1.1\times10^{-5}$ & $8\times10^{-6}$ & 0.998 & $-3.36\times10^{-7}$ & $1.09\times10^{-5}$ & $2.66\times10^{-5}$ & $-4.21\times10^{-5}$\\
    \hline
    \end{tabular}

\section{LEAVE-ONE-OUT CROSS-VALIDATION METRICS FOR THE EIGHT DIMENSIONAL GPR MODELS}
\label{app:LOO_8d}
    \captionof{table}{\justifying Shown are the root-mean-square error, mean absolute error, coefficient of determination, mean, standard deviation, maximum, and minimum.}
    \label{tab:LOO_8d}
    \setlength{\tabcolsep}{8.5pt}
    \begin{tabular}{cccccccc}
    \hline Quantity & RMSE & MAE & $R^2$ & Mean & SD & Max & Min \\
    \hline $\Delta\Omega_0$ & $1.3\times10^{-5}$ & $1\times10^{-5}$ & $0.995$ & $-3.86\times10^{-7}$ & $1.25\times10^{-5}$ & $5.41\times10^{-5}$ & $-4.596\times10^{-5}$ \\
    $\Delta\dot{a}_0$ & $1\times10^{-5}$ & $8\times10^{-6}$ & $0.996$ & $-7.90\times10^{-8}$ & $1.04\times10^{-5}$ & $6.45\times10^{-5}$ & $-3.97\times10^{-5}$ \\
   \hline
   \end{tabular}
\end{widetext}

\bibliographystyle{apsrev4-2}
\bibliography{references}

\end{document}